# Digital Natives, Digital Activists: Youth, Social Media and the Rise of Environmental Sustainability Movements

Manya Pandit[1] , Triveni Magadum[3,5] , Harshit Mittal[4,5,6], Omkar Kushwaha[4,7,8*]

[1]University School of Management Studies, Guru Gobind Singh Indraprastha University, Dwarka, Delhi-110078

[2]Doctor of Philosophy, Singhania University, Pacheri Bari, Jhunjhunu-333515, Rajasthan, India

[3]Computer Science Department, KLE Technological University, Hubballi-580031, Karnataka, India

[4]Center for Energy and Environment, School of Advanced Sciences, KLE Technological University, Hubballi-580031, Karnataka, India

[5]Pro H2Vis Solutions, KLE Technological University, Huballi-580031, Karnataka, India

[6]University School of Chemical Technology, Guru Gobind Singh Indraprastha University, Dwarka, Delhi-110078, India

[7]Energy Consortium, ICAR, Indian Institute of Technology, Madras, Chennai-600036, India

[8]Chemical Engineering Department, Indian Institute of Technology, Madras, Chennai-600036, India

*Corresponding email: Kushwaha.iitmadras@gmail.com

**Abstract**

The research examines the challenges revolving around young people's social movements, activism regarding sustainability, as well as the accompanying social media aspect, and how social media impacts environmental action. This study focuses on the environmental craze on social media platforms and its impact on young activists aged 16-25. With the advancement of social media, new avenues have opened for participation in sustainability issues, especially for the marginalized, as information moved through transnational networks at lightning speed. Along with specific Formative Visual Storytelling methods, the young leaders of the movement deploy hashtags and other online tools to capture the attention of their peers and decision makers. Challenges persist with "clicktivism," fatigue from the internet, and site limitations. This article contributes to insights on emerging forms of civic activism by explaining how digital natives adapt technology to reframe green activism. The research suggests that effective digital environmental movements integrate online and offline action, make it simple for individuals to get involved, and promote tolerance to algorithmic modifications and climate care among participants.

**1. Introduction**

The coincidence of the most urgent global environmental challenges with the information revolution has created a singular opportunity for civic mobilization, and most especially, among the youth generations. With the effects of climate change increasingly manifested in terms of extreme weather phenomena, loss of biodiversity, and resource limitations, a generation growing up with information technology has answered back by initiating new modes of environmental activism. The focus of this study is on the ways young people use social media to formulate, disseminate, and participate in sustainability movements that transcend traditional forms of activism (Burkhardt et al., 2016; Foos et al., 2021; Heaney & Rojas, 2014; Milan, 2019; Murtagh, 2016).

Youths of today live with dual citizenship, operating at the same time in physical communities and online spaces that inform their identities, relationships, and citizenship. Social media platforms are not only tools of communication but infrastructures per se through which, for the first generation of authentic "digital natives," they experience and engage with issues in society. Environmental sustainability activism has especially prospered in this type of internet environment because teenagers take advantage of the networked, participatory, and pictorial aspects of sites such as Instagram, TikTok, and Twitter to campaign their peers on environmental issues. The rate at which social networking transmits material has made possible the spread of youth environmentalism at a speedy rate, evading institutional barriers and gatekeepers that had disqualified youth from the environmental conversation until now (Cortés-Ramos et al., 2021; Earl et al., 2017; Hruska & Maresova, 2020; Sarker et al., 2015; Weller, 2016).

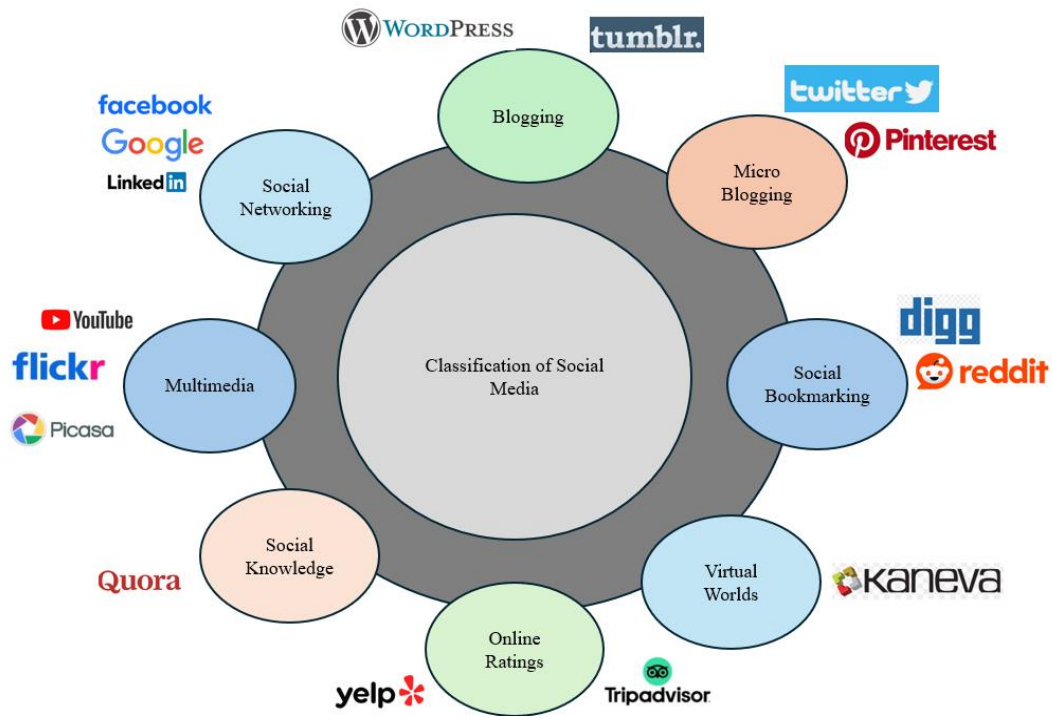

**Figure 1:** Classification of Social Media platforms.

Although increasing academic attention has been directed toward digital activism in general, comparatively few studies have directly explored how youth environmental sustainability movements function within and between social media environments. Such an omission matters especially because young people's views of environmental matters tend to differ from those of previous generations, both in terms of defining problems and in suggested remedies (Dirzo & Raven, 2003; Ellis et al., 2012; Forester & Machlist, 1996; Gibbs, 2000). Understanding how youth mobilize around environmental concerns to create an online movement and how those online movements, in turn, influence offline behaviors, policies, and systems is crucial information for environmental communicators, policymakers, educators, and platform builders. This study, therefore, investigates the specific strategies, obstacles, and impacts of youth environmental activism online, as well as specifically how young people themselves perceive the relationship between their online activism and environmental change offline (Mei, 2021; Mittal & Kushwaha, 2024; O'Connor et al., 2002; Pearce et al., 2019; Rose, 1990).

## 2. Methodology

The mixed-method approach was applied to this study in investigating online youth environmental activism. Sensitive to the biased interaction between online messages and offline results, the research design applied quantitative data gathering for supporting grand patterns and qualitative procedures in looking for youths' meaning-making and strategic reasoning among environmental activists. Methodological orientation was premised on digital ethnographic sensitivities that situate social media spaces as bounded cultural worlds that possess distinct norms, languages, and modes of interaction. As part of such a method, young environmental campaigners are situated not as objects of study but rather as digital cultural makers working within emergent media worlds to champion causes of sustainability (Garcia et al., 2017; Halog et al., 2011; Mascarenhas et al., 2019).

The research design was developed through rigorous consultation with media studies scholars and youth environmental groups to ensure methodological nuance while being responsive to the rapidly evolving nature of digital activism. Particular attention was paid to developing methods that would be capable of capturing both the content of youth environmental messages and the networked environments in which they operate. Through the utilization of three or more sources of data, the research aimed to produce a general picture of how youth sustainability movements function in online environments and to take into consideration the constraints in studying changing online processes (Badler & Smoliar, 1979; Çöteli, 2019a, 2019b).

2.1 Digital Ethnography

A six-month online ethnography was undertaken across five of the largest social media platforms (Instagram, TikTok, Twitter, YouTube, and Facebook) to watch youth environmental activism in its natural habitat. Researchers engaged with youth environmental online spaces, tracking 150 youth activist handles and 75 environmental campaign hashtags while systematically recording content types, interaction trends, and communication practices used across platforms. This ethnographic approach provided crucial contextual information on the online environments in which young people's environmental activism occurs, gathering 1,200 pieces of original content on environmental activism created by users aged 16-25 and revealing platform-specific cultural norms and communication practices that are impossible to access otherwise (Jacqmarcq, 2021; Tim et al., 2018).

2.2 In-depth interviews

In-depth interviews with 35 young environmental activists (ages 16-25) who have active digital presences tied to sustainability campaigning purposively sampled to be diverse in terms of geographic locations, platform specializations, environmental issue focus, and following size. The interview guide addressed five thematic domains: individual trajectories to digital environmental activism, strategic choices regarding platform utilization, experiences of algorithmic system navigation, perceived connections between digital advocacy and offline effects, and difficulties of maintaining digital activism habits. Each interview utilized photo-elicitation methods based on participants' social media postings to elicit in-depth discussion, took 60-90 minutes, and was held via video conference, with follow-up interviews with 12 participants to probe developing themes more in-depth.

2.3 Content Analysis

Our in-depth content analysis analyzed 500 high-engagement youth environmental posts on five online platforms (Instagram, n=150; TikTok, n=125; Twitter, n=100; YouTube, n=75; Facebook, n=50), using an advanced coding framework built through an iterative research process. The method focused on a strong analytical framework with 42 variables strategically categorized into six essential categories: content features, technical features, mobilization strategies, narrative structures, scientific content, and engagement measures. Through a systematic analysis of these dimensions, we were trying to tap into the varied communication tactics utilised by younger environmental activists through various digital ecosystems. Ensuring methodological appropriateness, two researchers trained independently coded an exemplary 20% representation of the sample and cross-validated the data to ascertain high interpretive reliability levels as well as render rich descriptive accounts of environmental digital communication approaches (Daniell et al., 2002; Daniell & Daniell, 1999; Fabbri & Tronchin, 2015; Zhou et al., 2008).

The research recognized the unique communication dynamics of each social networking site and the recognition that environmental messaging in each online environment varies significantly. We were then able to soak up the platform-specific nuances that might affect youth activist communication styles via sample stratification. Six analysis categories enabled a multifaceted investigation of how adolescent activists produce and spread environmental messages, ranging from the technical act of message creation to the narrational methods through which they move the audience. Additionally, it offered an avenue of understanding the emerging digital environmentalism trend, marking the creative nature through which youngsters leverage technology as a tool in raising awareness, as well as calling for social change.

2.4 Platform Analysis

A systematic study of platform technical affordances was conducted to position youth activist strategies within the technological environments they are operating in, comparing platform-specific features, algorithmic bias, governance rules, and technical constraints relevant to environmental communication on the five platforms in question. The affordance analysis framework assessed seven aspects of platform architecture: options for content formats, factors of visibility controlled by algorithms, community formation mechanisms, monetization models, content moderation policies, data policies, and cross-platform integration capabilities. This technical environment was essential for understanding strategic choices by young activists and making sense of how environmental messaging was negotiated to leverage or circumvent platform-specific constraints, particularly as young activists increasingly had sophisticated understandings of the implications of platform design choices for environmental visibility and reach (Allen et al., 2022; Carlsson & Jacobsson, 1997; Harper & Snowden, 2017).

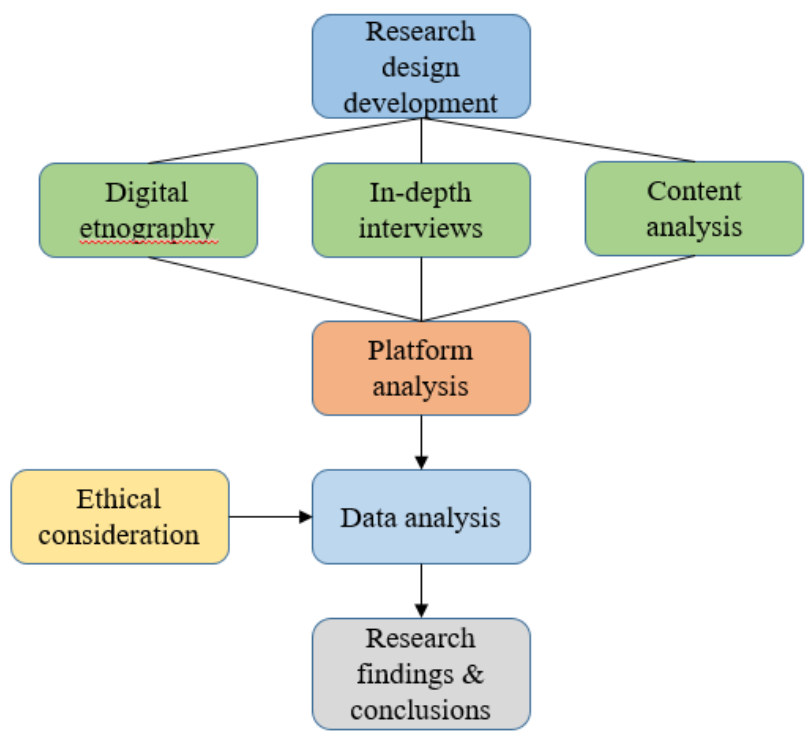

**Figure 2:** Mixed methods research process flowchart.

2.5 Data Analysis

The multi-modal data obtained were subjected to inductive and deductive analysis processes aimed at aligning emergent knowledge and theoretical ideas of environmental communication, youth studies, and digital media scholarship. The interview transcripts and ethnographic field notes were coded using iterative thematic coding, starting with open coding followed by axial coding to establish relationships between concepts. The task was facilitated by employing NVivo computer software to facilitate the massive qualitative dataset and aid in systematic comparison across participant reports. Content analysis data were scrutinized with comparative analysis and descriptive statistics by both platform and by type of issue, with reference specifically to variations in message framing and visual approach by environmental issue category. This numerical value was complemented with model post discourse analysis, with excellent engagement levels, to identify rhetorical tendencies that appealed to youth digital communities.

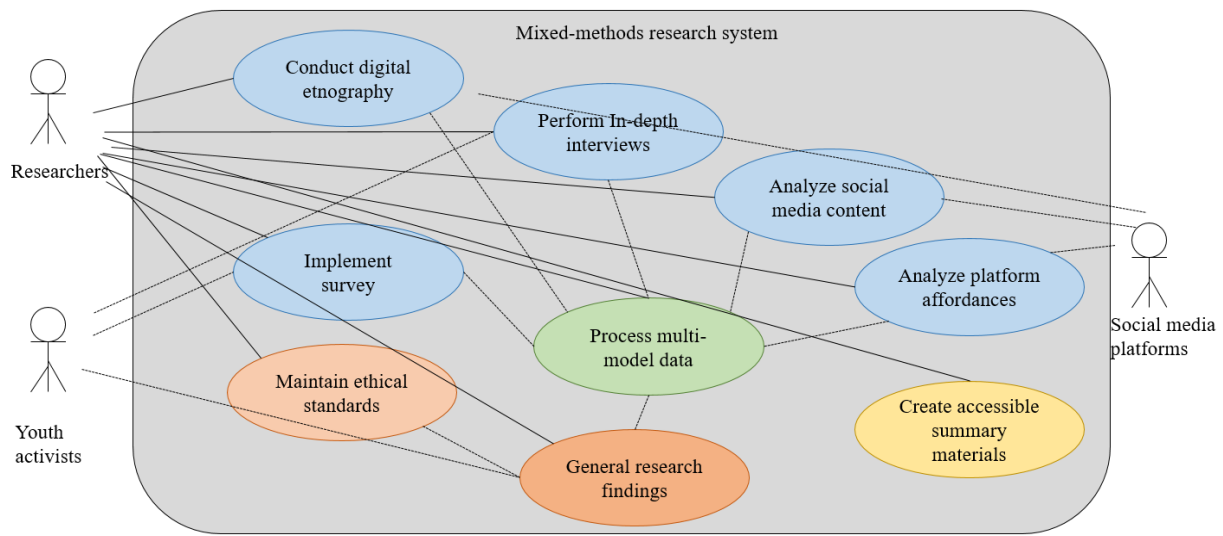

**Figure 3:** Research use case diagram.

Network analysis was used to follow and analyze relationships between different youth environmental campaigns and networks using hashtag data. NodeXL software was used to map and analyze patterns of relationships between institutional targets, participant accounts, and campaign hashtags to examine the structure of digital environmental movement networks. Through this, we found both centralized influencer hubs and distributed peer networks co-existing in youth environmental spaces. All sources of data were triangulated to reveal consistent patterns and interesting divergences, rigorously increasing findings' validity through methodological integration. Analytical memos tracking emergent theoretical understanding and methodological issues were kept by the research team throughout the process of analysis, informing later data collection waves and subsequent analysis refinements. Member checking was conducted with a subset of interview participants to check interpretations and analytical findings.

2.6 Ethical Considerations

The research plan was cleared by the Institutional Review Board of the university. Extra caution was exercised to protect the privacy and well-being of minors, with the consent of parents being sought from all participants younger than 18 years. Furthermore, as a result of the possible susceptibility of young environmental activists to harassment on the internet, all participant data were anonymized, and particular identification information of smaller accounts was masked in the final reporting. The research team established detailed protocols for coming across potentially distressing eco-anxiety or climate grief content, such as referral routes to access support resources for participants and researchers. As a token of appreciation for the efforts of the participants, the research team also committed to producing accessible summary materials for youth environmental groups as well as regular academic outputs.

**3. Results**

The study identified several key trends in the use of social media among youth for environmental sustainability campaigns, with both innovative approaches and constant challenges in activism online. The subsequent section offers the key findings organized under four broad themes found to be repeated across data sources. The combination of qualitative observations from participant interviews with quantitative trends from content analysis provides a rich picture of youth-driven digital environmental activism as it is expressed across current social media environments. These results add to new scholarship on digital civic action while directly engaging the distinctive intersection of youth identity construction, technological mediation, and environmental concern that defines current sustainability movements in digital environments.

3.1 Platform-Specific Engagement Tactics

Young environmental activists showed mastery of platform-specific affordances with a sophisticated ability to tailor their messaging and strategies to suit each platform. On visual platforms such as Instagram and TikTok, members focused on creating visually appealing content, conveying environmental ideas through understandable visual narratives. Content analysis showed that TikTok videos employing humor and popular formats gained 3.7 times higher engagement than plain informative content. On Twitter, young activists labored on rapid responses to environmental news headlines, direct conversation with institutional handles, and hashtag coordination for global climate events. Findings from interviews indicated that platform selection was not accidental but instead intentional, as activists were careful to select platforms based on audience demographics and communication goals. An interview respondent clarified: "I use Instagram to construct community through a visual narrative of local green action, but I use Twitter if I want to pressure a politician or a company, it's where you can create public accountability."

| Platform | Primary Engagement Strategies | Content Type Effectiveness | Activist-Reported Benefits | Key Challenges | Engagement Multiplier |
|---|---|---|---|---|---|
| Instagram | Visual storytelling, carousel educational content, sustainable lifestyle modeling | Aesthetically-driven content with actionable information $(3.2x)^2$ | Community building, visual documentation of environmental projects | Algorithm deprioritization of activist content | Medium (2.4x) |
| TikTok | Humor-based environmental education, trending format adaptation, challenge creation | Humor and trending formats $(3.7x)^2$, short-form educational content $(2.5x)^2$ | Rapid audience growth, youth peer influence, simplified complex concepts | Content depth limitations, ephemeral engagement | High (3.8x) |
| Twitter | Rapid response to environmental news, direct engagement with institutions, hashtag coordination | Real-time updates $(2.8x)^2$, institutional accountability threads $(3.3x)^2$ | Direct pressure on corporations/politicians, quick mobilization | Harassment of youth activists, context collapse | Medium (2.2x) |
| YouTube | In-depth environmental explainers, documentary-style content, collaborative videos | Long-form educational content $(1.9x)^2$, personal narrative + science $(2.7x)^2$ | Thorough concept exploration, monetization potential | High production expectations, slower growth | Low (1.7x) |
| Discord/ Private Communities | Private support networks, campaign coordination, detailed action planning | Resource sharing (N/A), campaign planning (N/A) | Safe spaces for emotional support, detailed planning without platform constraints | Limited public visibility, echo chamber potential | N/A³ |

**Table 1:** Cross-Platform comparison of youth environmental activism strategies and engagement metrics.

3.2 Digital Repertoires of Contention

The research identified five digital action repertoires of youth environmental movements. The first is information sharing, where complex environmental science and policy are translated into shareable, understandable information to promote ecological literacy among peers. The second is alternative representation, where counter-narratives to mainstream media reporting on environmental issues, in particular those of frontline communities and youth voices, systematically marginalized in traditional media, are created. Third is networked mobilization, organizing coordinated online-offline action in space, including virtual climate strikes, hashtag photo challenges, and decentralized direct action. Fourth is corporate and political pressure, coordinating collective action campaigns on particular institutions through coordinated messaging, boycotts, and corporate or political environmental failure amplification. The fifth repertoire is lifestyle advocacy, which showcases and encourages sustainable consumption and behavior decisions through personal experience, tutorials, and peer-to-peer influence networks. Content analysis illustrated that campaigns using multiple repertoires resulted in far greater engagement rates, with the most effective youth movements integrating online advocacy with offline action seamlessly.

3.3 Psychological Factors of Online Environmental Activism

One of the common themes running through the interviews was the insidious emotional terrain traversed by young online green activists. The respondents all spoke of significant psychological gains from activism online, ranging from diminished climate anxiety through collaborative action to the building of supportive communities and empowerment in countering environmental challenges. Yet many also spoke of experiences of digital exhaustion, overwhelm from repeated exposure to environmental crises, and frustration with perceived constraints of online activism. The evidence is of advanced coping mechanisms created by these long-term autistic participants in online environmental activism. These included intentional avoidance boundaries, incremental wins being marked, establishment of in-group digital support frameworks beyond public arenas of advocacy, and strategic disconnection phases. In the words of one 19-year-old participant: "You have to build what I call 'sustainable activism' habits of engagement that don't drain you because environmental issues are not going away after a single viral campaign."

3.4 Cross-Platform Movement Building

Content analysis and network visualization indicated the emergence of high-level cross-platform environmental campaigns crafted by youth activists. Rather than applying each platform as a discrete space, successful youth movements constructed unified media ecosystems with platform-specific content flowing into an overall campaign narrative. For instance, one campaign against fast fashion by youth used TikTok for viral challenges, Instagram for informative carousels regarding the environmental degradation of garment production, Twitter for calling out brands directly, and Discord for organizing and mobilizing communities. Interview evidence validated this cross-platform strategy as intentional, with the activists following a clear conscious process of building content strategies that took advantage of the unique capabilities of each platform while avoiding algorithm weaknesses. As platforms increasingly embraced algorithmic moderation of environmental (and particularly climate politics) content, young activists were highly adaptable, learning coded language, strategic tagging protocols, and platform-switching tactics to maintain movement momentum despite technical constraints.

## 4. Conclusion

This study shows that youth-led digital environmental activism is a recent evolution in civic engagement, characterized by particular strategies that leverage social media affordances while operating within platform limitations. The results prove that young green activists are neither naive techno-optimists nor simply engaging in "slacktivism" but rather employ digital technologies with a tactical purpose and have a clear-sighted recognition of the need for off-line systemic transformation. There are several important implications of this research. Second, the research dismisses oversimplified models of young digital activism by mapping their advanced movement-building strategies employed across platforms. Young climate movements have also developed new strategies for remaining in the spotlight, bridging online and offline organizing, and pushing back against algorithmic control and climate trauma. Second, the study points out the importance of platform structure and regulation in supporting or inhibiting environmental activism, which suggests that the technology companies have much to answer for in matters of citizens' democratic participation in environmental debate.

This research also uncovers persistent dilemmas confronting environmental movements online, including persistent contrasts between viral activism and policy outcomes, challenges of maintaining long-term participant involvement above trending periods, and the one-sided distribution of digital access and capabilities that can strengthen original inequality within environmental movements. Follow-up studies should look into how young people's online environmentalism develops over extended periods, with a particular focus on its institutional implications and the evolution of internal modes of governance. As environmental problems intensify and digital spaces continue to shift, youth-led sustainability movements at this intersection offer rich laboratories for teasing out new forms of civic participation. By adapting rapidly to platform changes while maintaining focus on core environmental values, these movements demonstrate how digital natives are reimagining activism for a connected world facing unprecedented ecological challenges. The most successful movements have effectively integrated digital and physical action, emotional support with strategic advocacy, and immediate campaign goals with long-term movement building, offering important models for environmental communication in the social media age.